\documentclass{PoS}

\newcommand{\url}[1]{\href{#1}{#1}}

\newcommand{\urlGammapySlack}{\href{https://gammapy.slack.com}{gammapy.slack.com}}
\newcommand{\urlCtaPipe}{\href{https://github.com/cta-observatory/ctapipe}{github.com/cta-observatory/ctapipe}}
\newcommand{\urlPint}{\href{https://github.com/nanograv/PINT}{github.com/nanograv/PINT}}
\newcommand{\urlFermipy}{\href{https://github.com/fermipy/fermipy}{github.com/fermipy/fermipy}}
\newcommand{\urlGammapyDocs}{\href{http://docs.gammapy.org}{docs.gammapy.org}}
\newcommand{\urlErfa}{\href{https://github.com/liberfa/erfa}{github.com/liberfa/erfa}}
\newcommand{\urlSofa}{\href{http://www.iausofa.org}{www.iausofa.org}}
\newcommand{\urlGammapyGithub}{\href{https://github.com/gammapy/gammapy}{github.com/gammapy/gammapy}}
\newcommand{\urlPytest}{\href{https://pytest.org}{pytest.org}}
\newcommand{\urlSphinx}{\href{http://www.sphinx-doc.org}{www.sphinx-doc.org}}
\newcommand{\urlJupyter}{\href{https://jupyter.org}{jupyter.org}}
\newcommand{\urlPypi}{\href{https://pypi.python.org}{pypi.python.org}}
\newcommand{\urlPip}{\href{https://pip.pypa.io}{pip.pypa.io}}
\newcommand{\urlAnacondaGammapy}{\href{https://anaconda.org/astropy/gammapy}{anaconda.org/astropy/gammapy}}
\newcommand{\urlGentooGammapy}{\href{https://packages.gentoo.org/packages/dev-python/gammapy}{packages.gentoo.org/packages/dev-python/gammapy}}
\newcommand{\urlGammapyForum}{\href{https://groups.google.com/forum/\#!forum/gammapy}{groups.google.com/forum/\#!forum/gammapy}}
\newcommand{\urlCtaAck}{\href{http://www.cta-observatory.org/consortium_acknowledgments}{www.cta-observatory.org/consortium\_acknowledgments}}
\newcommand{\urlGithub}{\href{https://github.com}{github.com}}
\newcommand{\urlRtd}{\href{https://readthedocs.org}{readthedocs.org}}
\newcommand{\urlTravis}{\href{https://travis-ci.org}{travis-ci.org}}
\newcommand{\urlAppveyor}{\href{https://appveyor.com}{appveyor.com}}
\newcommand{\urlSlack}{\href{https://slack.com}{slack.com}}

\newcommand{\urlHealpy}{\href{https://healpy.readthedocs.io}{healpy.readthedocs.io}}
\newcommand{\urlRegions}{\href{https://astropy-regions.readthedocs.io}{astropy-regions.readthedocs.io}}
\newcommand{\urlReproject}{\href{https://reproject.readthedocs.io}{reproject.readthedocs.io}}

\title{Gammapy -- A prototype for the CTA science tools}
\ShortTitle{Gammapy -- A prototype for the CTA science tools}

\author{
Christoph Deil$^a$,
Roberta Zanin$^a$,
\speaker{Julien Lefaucheur}$^b$,
Catherine Boisson$^b$,
Bruno Kh\'elifi$^c$,
R\'egis Terrier$^c$,
Matthew Wood$^d$,
Lars Mohrmann$^e$,
Nachiketa Chakraborty$^a$,
Jason Watson$^q$,
Rub\'en L\'opez Coto$^a$,
Stefan Klepser$^f$,
Matteo Cerruti$^g$,
Jean-Philippe Lenain$^g$,
Fabio Acero$^h$,
Arache Djannati-Ata{\"\i}$^c$,
Santiago Pita$^c$,
Zeljka Bosnjak$^i$,
Jos\'e Enrique Ruiz$^j$,
Cyril Trichard$^k$,
Thomas Vuillaume$^l$,
for the CTA Consortium,
Axel Donath$^a$,
Johannes King$^a$,
L\'ea Jouvin$^c$,
Ellis Owen$^m$,
Manuel Paz Arribas$^n$,
Brigitta Sipocz$^o$,
Dirk Lennarz$^p$,
Arjun Voruganti$^a$,
Marion Spir-Jacob$^c$
\\
\llap{$^a$}MPIK, Heidelberg, Germany\\
\llap{$^b$}LUTH, Obs. de Paris/Meudon, France\\
\llap{$^c$}APC/CNRS, Paris, France\\
\llap{$^d$}SLAC National Accelerator Laboratory, US\\
\llap{$^e$}FAU, Erlangen, Germany\\
\llap{$^f$}DESY, Zeuthen, Germany\\
\llap{$^g$}LPNHE, Paris, France\\
\llap{$^h$}CEA/IRFU, Saclay, France\\
\llap{$^i$}University of Rijeka, Croatia\\
\llap{$^j$}IAA-CSIC, Granada, Spain\\
\llap{$^k$}CPPM, Marseille, France\\
\llap{$^l$}LAPP, Annecy-le-Vieux, France\\
\llap{$^m$}UCL-MSSL, Dorking, United Kingdom\\
\llap{$^n$}Humboldt University, Berlin, Germany\\
\llap{$^o$}Cambridge, UK\\
\llap{$^p$}Georgia Tech, Atlanta, US\\
\llap{$^q$}University of Oxford, UK\\
E-mail:
\email{Christoph.Deil@mpi-hd.mpg.de},
\email{Roberta.Zanin@mpi-hd.mpg.de},
\email{julien.lefaucheur@obspm.fr},
\email{catherine.boisson@obspm.fr},
\email{khelifi@apc.in2p3.fr},
}

\abstract{

Gammapy is a Python package for high-level gamma-ray data analysis built on
Numpy, Scipy and Astropy. It enables us to analyze gamma-ray data and to create 
sky images, spectra and lightcurves, from event lists and instrument response 
information, and to determine the position, morphology
and spectra of gamma-ray sources.

So far Gammapy has mostly been used to analyze data from H.E.S.S. and Fermi-LAT,
and is now being used for the simulation and analysis of observations from the
Cherenkov Telescope Array (CTA). We have proposed Gammapy as a prototype for the
CTA science tools. This contribution gives an overview of the Gammapy package
and project and shows an analysis application example with simulated CTA data.

}

\FullConference{35th International Cosmic Ray Conference --- ICRC2017\\
		10--20 July, 2017\\
		Bexco, Busan, Korea}

\begin{document}

\section{Introduction}
\label{sec:intro}

% CTA and CTA science tools introduction
The Cherenkov Telescope Array (CTA) will observe the sky in very-high-energy
(VHE, E > 20$\,$GeV) gamma-ray light soon. CTA will consist of large telescope
arrays at two sites, one in the southern (Chile) and one in the northern (La
Palma) hemisphere. It will perform surveys of large parts of the sky, targeted
observations on Galactic and extra-galactic sources, and more specialized
analyses like a measurement of charged cosmic rays, constraints on the
intergalactic medium opacity for gamma-rays and a search for dark matter.
Compared to current Cherenkov telescope arrays such as H.E.S.S., VERITAS or
MAGIC, CTA will have a much improved detection area, angular and energy
resolution, improved signal/background classification and sensitivity. CTA is
expected to operate for thirty years, and all astronomers will have access to
CTA high-level data, as well as CTA science tools (ST) software. The ST can be
used for example to generate sky images and to measure source properties such as
morphology, spectra and light curves, using event lists as well as instrument
response function (IRF) and auxiliary information as input. 

\begin{figure}[t]
\centering
\includegraphics[width=0.5\textwidth]{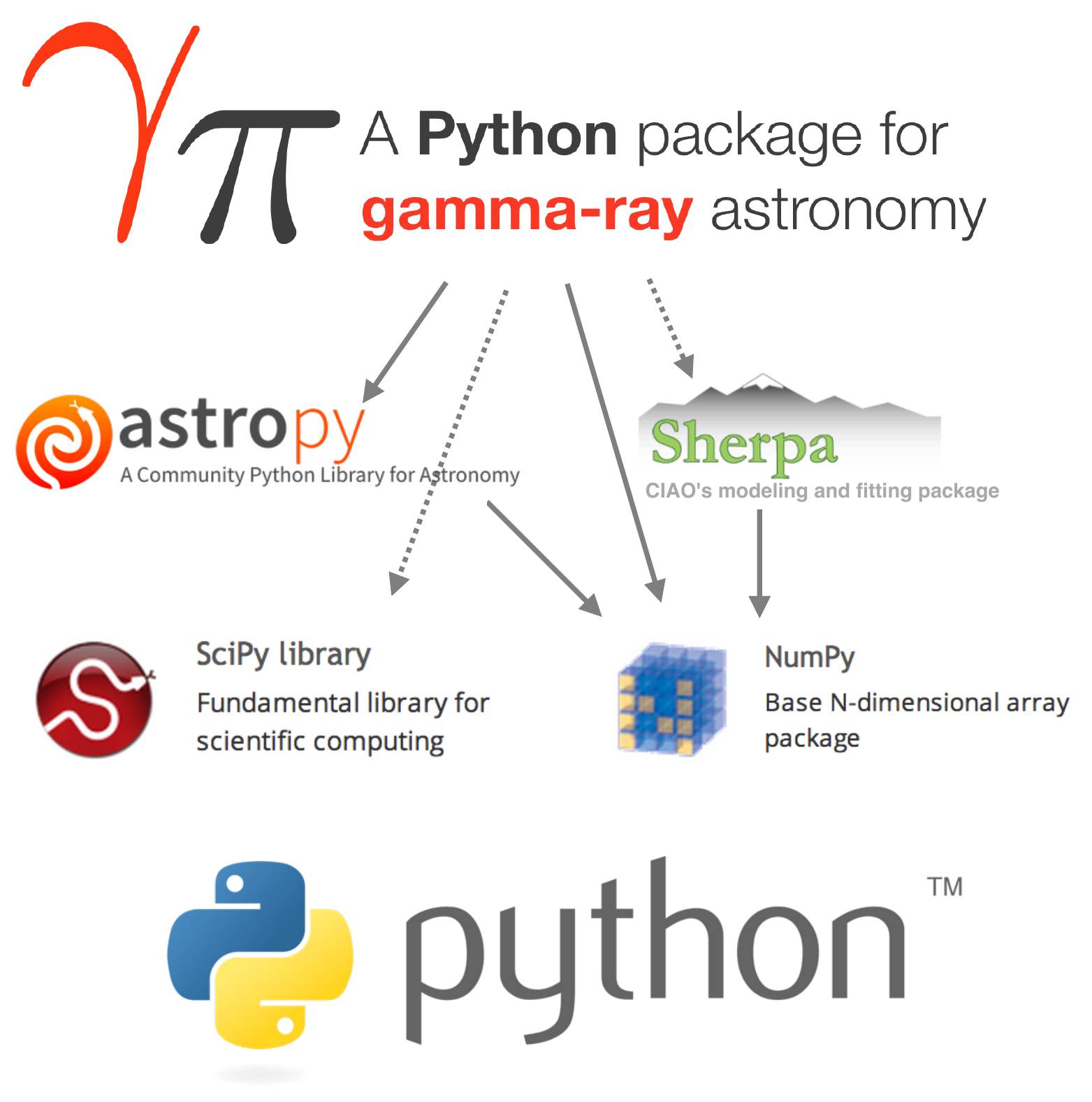}
\caption{
The Gammapy software stack. Required dependencies (Python, Numpy and Astropy)
are illustrated with solid arrows, optional dependencies (Scipy and Sherpa) with
dashed arrows.
}
\label{fig:stack}
\end{figure}

% What is Gammapy?
Gammapy is a prototype for the CTA ST, built on the scientific Python stack and
Astropy \cite{astropy}, optionally using Sherpa \cite{sherpa2001, sherpa2009,
sherpa2011} or other packages for modeling and fitting (see
Figure~\ref{fig:stack}). A 2\hbox{-}dimensional analysis of the sky images for
source detection and morphology fitting, followed by spectral analysis, was
first  implemented. A 3\hbox{-}dimensional analysis with a simultaneous spatial
and spectral model of the gamma-ray emission, as well as background (called
``cube analysis'' in the following) is under development.
% Existing studies using Gammapy
A first study comparing spectra obtained with the classical 1D analysis and the
3D cube analysis using point source observations with H.E.S.S. is presented in
\cite{lea}. Further developments and verification using data from existing
Cherenkov telescope arrays such as H.E.S.S. and MAGIC, as well as simulated CTA
data is ongoing.
% Gammapy use in CTA
Gammapy is now used for scientific studies with existing ground-based gamma-ray
telescopes \cite{hgps, shells}, the Fermi-LAT space telescope \cite{owen2015},
as well as for CTA \cite{julien, roberta, cyril}.

% Outline of this paper
In this writeup we focus on the software and technical aspects of Gammapy.We
start with a brief overview of the context in Section~\ref{sec:context},
followed by a description of the Gammapy package in Section~\ref{sec:package},
the Gammapy project in Section~\ref{sec:project} and finally our conclusions
concerning Gammapy as as CTA science tool prototype in
Section~\ref{sec:conclusions}.

\section{Context}
\label{sec:context}

% Python, Numpy, Astropy
Prior to Gammapy, in 2011/2012, a collection of Python scripts for the analysis
of IACT data was released as a first test of open-source VHE data analysis
software: ``PyFACT: Python and FITS Analysis for Cherenkov Telescopes''
\cite{pyfact}. This project is not updated anymore, but it was the first
implementation of our idea to build a CTA science tools package based on  Python
and using Numpy and Scipy, during the first development stages of the CTA
project. In 2011, the astronomical Python community came together and created
the Astropy project and package \cite{astropy}, which is a key factor making
Python the most popular language for astronomical research codes (at least
according to this informal analysis \cite{perry} and survey
\cite{momcheva2015}). Gammapy is an Astropy affiliated package, which means that
where possible it uses the Astropy core package instead of duplicating its
functionality, as well as having a certain quality standard such as having
automated tests and documentation for the available functionality. In recent
years, several other packages have adopted the same approach, to build on
Python, Numpy and Astropy. To name just a few, there is ctapipe (\urlCtaPipe),
the prototype for the low-level CTA data processing pipeline (up to the creation
of lists of events and IRFs); Naima for modeling the non-thermal spectral energy
distribution of astrophysical sources \cite{naima}; PINT, a new software for
high-precision pulsar timing (\urlPint), and Fermipy (\urlFermipy), a Python
package that facilitates analysis of data from the Large Area Telescope (LAT)
with the Fermi Science Tools and adds some extra functionality.

We note that many other astronomy projects have chosen Python and Astropy as the
basis both for their data calibration and reduction pipelines and their science
tools. Some prominent examples are the Hubble space telescope (HST)
\cite{hubble}, the upcoming James Webb Space Telescope (JWST) \cite{jwst} and
the Chandra X-ray observatory \cite{sherpa2001, chandra}. Even projects like
LSST that started their analysis software developments before Astropy existed
and are based on C++/SWIG are now actively working towards making their software
interoperable with Numpy and Astropy, to avoid duplication of code and
development efforts, but also to reduce the learning curve for their science
tool software (since many astronomers already are using Python, Numpy and
Astropy) \cite{lsst}.

% Open data formats
For current ground-based IACTs, data and software are mostly private to the
collaborations operating the telescopes. CTA will be, for the first time in VHE
gamma-ray astronomy, operated as an open observatory. This implies fundamentally
different requirements for the data formats and software tools. Along this path,
the current experiments, H.E.S.S., MAGIC and VERITAS, have started converting
their data to FITS format, yet relying on different (some private, some open)
analysis tools, and many slightly different ways to store the same information
in FITS files appeared. The CTA high-level data model and data format
specifications  are currently being written and will give a framework to the
current experiments to store data. The methods to link events to IRFs still have
to be extended to  support multiple event types and the IRF and background model
formats will have to be  extended to be more precise \cite{opendata}. The
Gammapy team is participating  in and contributing to the effort to prototype
and find a good high-level data  model and formats for CTA.

\section{Gammapy package}
\label{sec:package}

Gammapy offers the high-level analysis tools to generate science results
(images, spectra, light curves and source catalogs) based on input data
consisting of reconstructed events (with an arrival direction and an energy)
that are  classified according to their types (e.g. gamma-like,
cosmic-ray-like). In the data processing chain of CTA, the reconstruction and
classification of events are realized by another Python package,
ctapipe\footnote{\urlCtaPipe}. Data are stored in FITS format. The high-level
analysis consists of:
\vspace{-0.3cm}
\begin{itemize}
\setlength\itemsep{-0.5em}

\item selection of a data cube (energy and positions) around a sky position from
all event lists,

\item computation of the corresponding exposure,

\item estimation the background directly from the data (e.g. with a
ring background model \cite{berge}) or from a model (e.g. templates built from
real data beforehand),

\item creation of sky images (signal, background, significance, etc) and
morphology fitting,

\item spectrum measurement with a 1D analysis or with a 3D analysis by adjusting both spectral and spatial shape of gamma-ray sources,

\item computation of light-curves and phasograms, search for transient signals,

\item derivation of a catalog of excess peaks (or a source catalog).

\end{itemize}
\vspace{-0.28cm}
For such an analysis, Gammapy is using the IRFs produced by the ctapipe  package
and processes them precisely to get an accurate estimation of high-level
astrophysical quantities.

\begin{figure}[t]
\centering
\includegraphics[width=0.7\textwidth]{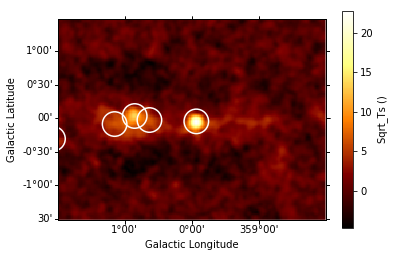}
\caption{
Application example: significance image for the Galactic centre region using
1.5~hours of simulated CTA data. White circles are peaks above 5~sigma.
}
\label{fig:app}
\end{figure}

The Gammapy code base is structured into several sub-packages dedicated to
specific classes  where each of the packages bundle corresponding functionality
in a namespace (e.g. data and observation handling in \textit{gammapy.data}, IRF
functionality in \textit{gammapy.irf}, spectrum estimation and modeling in
\textit{gammapy.spectrum} \ldots). The Gammapy features are described in detail
in the Gammapy documentation (\urlGammapyDocs) and many examples given in the
tutorial-style Jupyter notebooks, as well as in \cite{gammapy-icrc2015}.
Figure~\ref{fig:app} shows one result of the ``CTA data analysis with Gammapy''
notebook: a significance sky image of the Galactic center region using 1.5~hours
of simulated CTA data. The background was estimated using the ring background
estimation technique, and peaks above 5~sigma are shown with white circles. For
examples of CTA science studies using Gammapy, we refer you to other posters
presented at this conference: Galactic survey \cite{roberta}, PeVatrons
\cite{cyril} and extra-galactic sources \cite{julien}. Several other examples
using real data from H.E.S.S. and Fermi-LAT, as well as simulated data for CTA
can be found via \urlGammapyDocs\ by following the link to ``tutorial
notebooks''.
 
The Gammapy Python package is primarily built on Numpy \cite{numpy}, and Astropy
\cite{astropy} as core dependencies. Data is stored in Numpy arrays or objects
such as \textit{astropy.coordinates.SkyCoord} or \textit{astropy.table.Table}
that hold Numpy array data members. Numpy provides many functions for
array-oriented computing and numerics, and Astropy provides astronomy-specific
functionality. The Astropy functionality most commonly used in Gammapy is
\textit{astropy.io.fits} for FITS data I/O, \textit{astropy.table.Table} as a
container for tabular data (e.g. event lists, but also many other things like
spectral points or source catalogs), \textit{astropy.wcs.WCS} for world
coordinate systems mapping pixel to sky coordinates, as well as
\textit{astropy.coordinates.SkyCoord} and \textit{astropy.time.Time} objects to
represent sky coordinates and times. Astropy.coordinates  as well as
astropy.time are built on ERFA (\urlErfa), the open-source variant of this IAU
Standards of Fundamental Astronomy (SOFA) C library (\urlSofa). In Gammapy, we
use \textit{astropy.units.Quantity} objects extensively, where a quantity is a
Numpy array with a unit attached, supporting arithmetic in computations and
making it easier to read and write code that does computations involving
physical quantities.

\begin{figure}[t]
\centering
\includegraphics[width=0.5\textwidth]{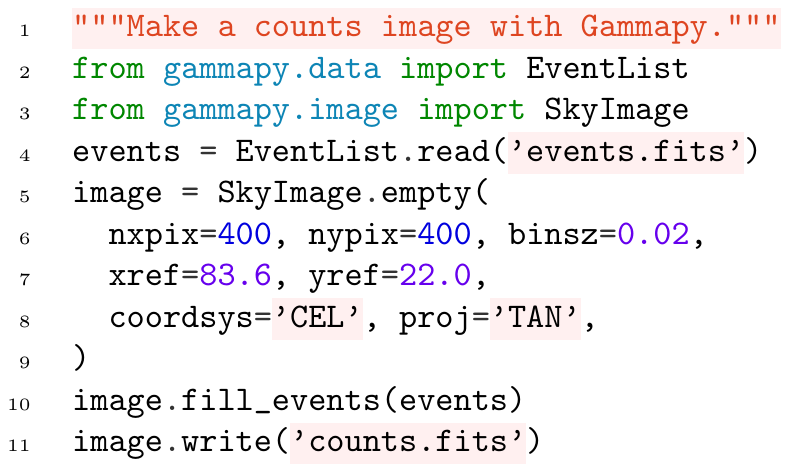}
\caption{
An example script using Gammapy to make a counts image from an event list. This
is used in Section~\protect\ref{sec:package} to explain how Gammapy achieves
efficient processing of event and pixel data from Python: all data is stored in
Numpy arrays and passed to existing C extensions in Numpy and Astropy.
}
\label{fig:code_example}
\end{figure}

As an example, a script that generates a counts image from an event list using
Gammapy is shown in Figure~\ref{fig:code_example}. The point we want to make
here is that it is possible to efficiently work with events and pixels and to
implement algorithms from Python, by storing all data in Numpy arrays and
processing via calls into existing C extensions in Numpy and Astropy. E.g. here
\textit{EventList} stores the RA and DEC columns from the event list as Numpy
arrays, and \textit{SkyImage} the pixel data as well, and
\textit{image.fill(events)}, and all processing happens in existing C extensions
implemented or wrapped in Numpy and Astropy. In this example, the \textit{read}
and \textit{write} methods call into \textit{astropy.io.fits} which calls into
CFITSIO (\cite{cfitsio}), and the \textit{image.fill(events)} method calls into
\textit{astropy.wcs.WCS} and WCSLib (\cite{wcslib}) as well as
\textit{numpy.histogramdd}. 

Gammapy aims to be a base package on which other more specialized packages such
as Fermipy (\urlFermipy) for Fermi-LAT data analysis or Naima \cite{naima} for
the modeling of non-thermal spectral energy distributions of astrophysical
sources can build. For this reason we avoid introducing new required
dependencies besides Numpy and Astropy. That said, Gammapy does import the
following optional dependencies to provide extra functionality (sorted in the
order of how common their use is within Gammapy). Scipy \cite{scipy} is used for
integration and interpolation, Matplotlib \cite{matplotlib} for plotting and
Sherpa \cite{sherpa2001, sherpa2009, sherpa2011} for modeling and fitting. In
addition, the following packages are used at the moment for functionality that
we expect to become available in the Astropy core package within the next year:
\textit{regions} (\urlRegions) to handle sky and pixel regions,
\textit{reproject} (\urlReproject) for reprojecting sky images and cubes and
\textit{healpy} (\urlHealpy) for HEALPix data handling.

\section{Gammapy project}
\label{sec:project}

In this section we describe the current setup of the Gammapy project. We are using the common tools and services for Python open-source projects for software
development, code review, testing, documentation, package distribution and user
support.

Gammapy is distributed and installed in the usual way for Python packages. Each
stable release is uploaded to the Python package index (\urlPypi), and
downloaded and installed by users via \textit{pip install gammapy} (\urlPip).
Binary packages for conda are available via the conda Astropy channel
(\urlAnacondaGammapy) for Linux, Mac and Windows, which conda users can install
via \textit{conda install gammapy -c astropy}. Binary packages for the Macports
package manager are also available, which users can install via \textit{port
install gammapy}. At this time, Gammapy is also available as a Gentoo Linux
package (\urlGentooGammapy) and a Debian Linux package is in preparation.

Gammapy development happens on Github (\urlGammapyGithub). We make extensive use
of the pull request system to discuss and review code contributions. For testing
we use pytest (\urlPytest), for continuous integration Travis-CI (Linux and Mac)
as well as Appveyor (Windows). For documentation Sphinx (\urlSphinx), for
tutorial-style documentation Jupyter notebooks (\urlJupyter) are used. For
Gammapy developer team communication we use Slack (\urlGammapySlack). A public
mailing list for user support and discussion is available (\urlGammapyForum).
Two face-to-face meetings for Gammapy were organized so far, the first on in
June 2016 in Heidelberg as a coding sprint for developers only, the second on in
February 2017 in Paris as a workshop for both Gammapy users and developers.

Gammapy is under very active development, especially in the area of modeling,
and in the implementation of a simple-to-use, high-level end-user interface
(either config file driven or command line tools). We will write a paper on
Gammapy later this year and are working towards a version 1.0 release.

\section{Conclusions}
\label{sec:conclusions}

In the past two years, we have developed Gammapy as an open-source analysis
package for existing gamma-ray telescope and as a prototype for the CTA science
tools. Gammapy is a Python package, consisting of functions and classes that can
be used as a flexible and extensible toolbox to implement and execute high-level
gamma-ray data analyses.

We find that the Gammapy approach, to build on the powerful and well-tested
Python packages Numpy and Astropy, brings large benefits: a small codebase that
is focused on gamma-ray astronomy in a single high-level language is easy to
understand and maintain. It is also easy to modify and extend as new use cases
arise, which is important for CTA, since it can be expected that the modeling of
the instrument, background and astrophysical emission, as well as the analysis
method in general (e.g. likelihood or Bayesian statistical methods) will evolve
and improve over the next decade. Last but not least, the Gammapy approach is
inherently collaborative (contributions from $\sim$30 gamma-ray astronomers so
far), sharing development effort as well as know-how with the larger
astronomical community, that to a large degree already has adopted Numpy and
Astropy as the basis for astronomical analysis codes in the past 5~years.

\section{Acknowledgements} \label{sed:acknowledgements}

This work was conducted in the context of the CTA Consortium. We gratefully
acknowledge financial support from the agencies and organizations listed here:\\
\urlCtaAck

We would like to thank the Scientific Python and specifically the Astropy
community for providing their packages which are invaluable to the development
of Gammapy, as well as tools and help with package setup and continuous
integration, as well as building of conda packages.

We thank the GitHub (\urlGithub) team for providing us with an excellent free
development platform, ReadTheDocs (\urlRtd) for free documentation hosting,
Travis (\urlTravis) and Appveyor (\urlAppveyor) for free continuous integration
testing, and Slack (\urlSlack) for a free team communication channel.

\bibliography{gammapy-icrc2017}

\providecommand{\href}[2]{#2}\begingroup\raggedright\begin{thebibliography}{10}

\bibitem{astropy}
{Astropy Collaboration}, T.~P. {Robitaille}, E.~J. {Tollerud} and
  P.~{Greenfield et al.}, \emph{{Astropy: A community Python package for
  astronomy}}, \href{https://doi.org/10.1051/0004-6361/201322068}{\emph{\aap}
  {\bfseries 558} (Oct., 2013) A33}.

\bibitem{sherpa2001}
P.~{Freeman}, S.~{Doe} and A.~{Siemiginowska}, \emph{{Sherpa: a
  mission-independent data analysis application}},  in \emph{Astronomical Data
  Analysis}, vol.~4477 of \emph{\procspie}, pp.~76--87, Nov., 2001,
  \href{https://arxiv.org/abs/astro-ph/0108426}{{\ttfamily astro-ph/0108426}}.

\bibitem{sherpa2009}
B.~{Refsdal et al.}, \emph{{Sherpa: 1D/2D modeling and fitting in Python}},  in
  \emph{Proceedings of the 8th Python in Science Conference}, (Pasadena, CA
  USA), pp.~51 -- 57, 2009.

\bibitem{sherpa2011}
B.~Refsdal, S.~Doe, D.~Nguyen and A.~Siemiginowska, \emph{{Fitting and
  Estimating Parameter Confidence Limits with Sherpa}},  in \emph{10th SciPy
  Conference}, pp.~4 -- 10, 2011.

\bibitem{lea}
L.~{Jouvin et al.}, \emph{{Toward a 3D analysis in Cerenkov gamma-ray
  astronomy}},  in \emph{\textit{these proceedings}}, 2017.

\bibitem{hgps}
S.~{Carrigan et al. for the H.E.S.S. collaboration}, \emph{{The H.E.S.S.
  Galactic Plane Survey - maps, source catalog and source population}},
  {\emph{ArXiv e-prints} (July, 2013) },
  [\href{https://arxiv.org/abs/1307.4690}{{\ttfamily 1307.4690}}].

\bibitem{shells}
G.~{Puehlhofer et al. for the H.E.S.S. collaboration}, \emph{{Search for new
  supernova remnant shells in the Galactic plane with H.E.S.S.}},  vol.~34 of
  \emph{ICRC}, p.~886, July, 2015,
  \href{https://arxiv.org/abs/1509.03872}{{\ttfamily 1509.03872}}.

\bibitem{owen2015}
E.~{Owen}, C.~{Deil}, A.~{Donath} and R.~{Terrier}, \emph{{The gamma-ray Milky
  Way above 10 GeV: Distinguishing Sources from Diffuse Emission}},
  {\emph{ArXiv e-prints} (June, 2015) },
  [\href{https://arxiv.org/abs/1506.02319}{{\ttfamily 1506.02319}}].

\bibitem{julien}
J.~{Lefaucheur for the CTA consortium}, \emph{{Gammapy: high level data
  analysis for extragalactic science cases with the Cherenkov Telescope
  Array}},  in \emph{\textit{these proceedings}}, 2017.

\bibitem{roberta}
R.~{Zanin for the CTA consortium}, \emph{{Observing the Galactic Plane with
  Cherenkov Telescope Array}},  in \emph{\textit{these proceedings}}, 2017.

\bibitem{cyril}
C.~{Trichard for the CTA consortium}, \emph{{Searching for PeVatrons in the CTA
  Galactic Plane Survey}},  in \emph{\textit{these proceedings}}, 2017.

\bibitem{pyfact}
M.~{Raue} and C.~{Deil}, \emph{{PyFACT: Python and FITS analysis for Cherenkov
  telescopes}},  vol.~1505 of \emph{American Institute of Physics Conference
  Series}, pp.~789--792, Dec., 2012,
  \href{https://doi.org/10.1063/1.4772378}{DOI}.

\bibitem{perry}
P.~{Greenfield}, \emph{{What Python Can Do for Astronomy}},  vol.~442 of
  \emph{Astronomical Society of the Pacific Conference Series}, pp.~425--+,
  July, 2011.

\bibitem{momcheva2015}
I.~{Momcheva} and E.~{Tollerud}, \emph{{Software Use in Astronomy: an Informal
  Survey}}, {\emph{ArXiv e-prints} (July, 2015) },
  [\href{https://arxiv.org/abs/1507.03989}{{\ttfamily 1507.03989}}].

\bibitem{naima}
V.~{Zabalza}, \emph{{naima: a Python package for inference of relativistic
  particle energy distributions from observed nonthermal spectra}},
  {\emph{ArXiv e-prints} (Sept., 2015) },
  [\href{https://arxiv.org/abs/1509.03319}{{\ttfamily 1509.03319}}].

\bibitem{hubble}
P.~{Greenfield} and R.~L. {White}, \emph{{Where Will PyRAF Lead Us? The Future
  of Data Analysis Software at STScI}},  p.~437, Jan., 2006.

\bibitem{jwst}
H.~{Bushouse et al.}, \emph{{ {T}he {J}ames {W}ebb {S}pace {T}elescope {D}ata
  {C}alibration {P}ipeline }},  in \emph{{P}roceedings of the 14th {P}ython in
  {S}cience {C}onference}, pp.~44 -- 48, 2015.

\bibitem{chandra}
T.~{Aldcroft}, \emph{{Keeping the Chandra Satellite Cool with Python}},  in
  \emph{Proceedings of the 9th Python in Science Conference}, pp.~30 -- 34,
  2010.

\bibitem{lsst}
T.~{Jenness et al.}, \emph{{Investigating interoperability of the LSST data
  management software stack with Astropy}},  vol.~9913, pp.~99130G--99130G--13,
  2016, \href{https://doi.org/10.1117/12.2231313}{DOI}.

\bibitem{opendata}
C.~{Deil} and C.~{Boisson et al.}, \emph{{Open high-level data formats and
  software for gamma-ray astronomy}}, {\emph{ArXiv e-prints} (Oct., 2016) },
  [\href{https://arxiv.org/abs/1610.01884}{{\ttfamily 1610.01884}}].

\bibitem{berge}
D.~{Berge}, S.~{Funk} and J.~{Hinton}, \emph{{Background modelling in
  very-high-energy {$\gamma$}-ray astronomy}},
  \href{https://doi.org/10.1051/0004-6361:20066674}{\emph{\aap} {\bfseries 466}
  (May, 2007) 1219--1229},
  [\href{https://arxiv.org/abs/astro-ph/0610959}{{\ttfamily
  astro-ph/0610959}}].

\bibitem{gammapy-icrc2015}
A.~{Donath et al.}, \emph{{Gammapy - A Python package for gamma-ray
  astronomy}}, {\emph{ArXiv e-prints} (Sept., 2015) },
  [\href{https://arxiv.org/abs/1509.07408}{{\ttfamily 1509.07408}}].

\bibitem{numpy}
S.~Van Der~Walt, S.~C. Colbert and G.~Varoquaux, \emph{{The NumPy array: a
  structure for efficient numerical computation}}, {\emph{Computing in Science
  \& Engineering} {\bfseries 13} (2011) 22--30}.

\bibitem{cfitsio}
W.~D. {Pence}, ``{CFITSIO: A FITS File Subroutine Library}.'' Astrophysics
  Source Code Library, Oct., 2010.

\bibitem{wcslib}
M.~R. {Calabretta}, ``{Wcslib and Pgsbox}.'' Astrophysics Source Code Library,
  Aug., 2011.

\bibitem{scipy}
\emph{{{SciPy}: Open source scientific tools for Python}},  2001.

\bibitem{matplotlib}
J.~D. Hunter, \emph{{Matplotlib: A 2D graphics environment}}, {\emph{Computing
  In Science \& Engineering} {\bfseries 9} (2007) 90--95}.

\end{thebibliography}\endgroup
\bibliographystyle{JHEP}

\end{document}